\documentclass[prl,aps,singlecolumn,superscriptaddress,preprintnumbers,nopacs,floatfix,amsmath,amssymb]{revtex4}

\usepackage{graphicx}
\usepackage{dcolumn}
\usepackage{bm}

\usepackage{amsmath}
\usepackage{graphicx}
\usepackage{amssymb}
\usepackage{mathrsfs}
\usepackage{bbm}
\usepackage{epsfig}

\newcommand{\be}{\begin{eqnarray}}
\newcommand{\ee}{\end{eqnarray}}

\begin{document}

%
\title{ Covalent bond symmetry breaking and protein secondary structure
}

\author{Martin Lundgren}
\email{Martin.Lundgren@physics.uu.se}
\affiliation{Department of Physics and Astronomy, Uppsala University,
P.O. Box 803, S-75108, Uppsala, Sweden}
\author{Antti J. Niemi}
\email{Antti.Niemi@physics.uu.se}
\affiliation{
Laboratoire de Mathematiques et Physique Theorique
CNRS UMR 6083, F\'ed\'eration Denis Poisson, Universit\'e de Tours,
Parc de Grandmont, F37200, Tours, France}
\affiliation{Department of Physics and Astronomy, Uppsala University,
P.O. Box 803, S-75108, Uppsala, Sweden}

\begin{abstract}
\noindent
Both symmetry and organized  breaking of symmetry have a pivotal r\^ole in our understanding of 
structure and pattern formation   
in physical systems, including  the origin of mass in the Universe  and the chiral structure  of biological macromolecules.    
Here we report  on a new symmetry breaking phenomenon that takes place in all  biologically active proteins,
thus this symmetry breaking relates to  the  inception of life.   The unbroken symmetry 
determines  the covalent bond geometry of a sp3 hybridized carbon atom. It  dictates the 
tetrahedral architecture of atoms around  the central carbon of an amino acid.    
Here we show that in a biologically active protein this symmetry becomes broken.  
Moreover, we show that the pattern of symmetry
breaking  is in a direct correspondence with the  local secondary structure of the folded protein. 
\end{abstract}



\maketitle

Protein modeling  is based on various well tested and 
broadly accepted stereochemical paradigms  \cite{mol},  \cite{pro}. These paradigms  
are instrumental in protein structure prediction  \cite{ros}, and  underlie the
phenomenological  force fields that  describe protein  dynamics \cite{nat}. 
The enormous success  in  resolving  over  70.000 structures  
that are presently in Protein Data Bank (PDB) \cite{struc} is a clear manifestation that  the various 
paradigms are valid to a high precision. 
Among the  important paradigms  is the assumption  that the backbone C$_\alpha$ carbons are in a 
definite sp3 
hybridized state, with its distinct  tetrahedral geometry. 
For example the  backbone $\tau_{NC} \equiv$ (N-C$_\alpha$-C) bond angle
should always  fluctuate around a definite  and computable value
that  only depends on the covalent bonds  between the  C$_\alpha$ and its 
adjacent C$_\beta$, N, C and H atoms.  In particular, this value 
should {\it not}  depend on the secondary structure 
environment. 

With  the arrival of third-generation synchrotron  X-ray
sources and the ensuing rapid increase in the number of protein structures  that are being resolved 
with an ultrahigh sub-\.Angstr\"om resolution  it is now possible to experimentally scrutinize the validity of
these paradigms.
In particular any systematic, secondary structure dependent  {\it breaking} of the covalent tetrahedral 
symmetry around the C$_\alpha$ carbons   could help  us to better understand why proteins 
fold  and to predict  more accurately how they fold. This could also have major repercussions to pharmaceutical 
drug development, and to help us better understand what is life. 

A molecular dynamics force field explicitely assumes that the tetrahedral symmetry remains unbroken. But 
{\it ab initio} quantum mechanical calculations \cite{sch} and empirical 
studies \cite{krp1}-\cite{tou} have  already pointed out  that  tetrahedral bond angles around a sp3 
hybridized carbon may be subject to measurable fluctuations.
For example,  there is an estimate that  the $\tau_{NC} \equiv$ (N-C$_\alpha$-C) bond angle could
fluctuate as much as 8.8$^o$ \cite{krp1} around its equilibrium position. This would have clearly 
measurable effects on the way how proteins fold. But the potential existence 
of a systematic and secondary structure  dependent tetrahedral 
symmetry breaking have until now not been scrutinized.

Here we address the presence of a systematic tetrahedral 
symmetry breaking by  investigating  the secondary structure 
dependence in  the values of $\tau_{NC}$,  and in
the adjacent  $\tau_{N\beta} \equiv$ (N-C$_\alpha$-C$_\beta$)  and  $\tau_{C\beta}\equiv$ (C-C$_\alpha$-C$_\beta$)  
bond angles. In order to diminish any bias towards paradigm based refinements  
we inspect several subsets in Protein Data Bank (PDB).
These include the canonical one that comprises all  PDB 
configurations with resolution 2.0 \.A or better, and its subsets with resolution better than 1.5 \.A, and better than 1.0 \.A. We
also inspect  a subset of  the 2.0 \.A  set that contains only those  proteins that have less than 30$\%$ sequence similarity,  and finally we also consider those proteins that appear in the CATH classification. We find that our conclusions are independent of the data set we use, and for illustrative purposes we use the canonical 2.0\.A set. 

The conventional backbone Ramachandran torsion angles $\phi$, $\psi$ and $\omega$ relate to the
backbone atoms N and C that we  investigate. To diminish potential bias that may depend on refinement 
procedures, we here adopt the N and C independent, geometrically determined backbone Frenet frames;
we follow the construction described in  \cite{dff}.  
These frames  depend {\it only} on the positions of  the C$_\alpha$ carbon  coordinates ${\bf r}_i$ 
with $i=1,...,n$ labeling the residues. We first introduce the unit backbone tangent ($\mathbf t$) and binormal ($\mathbf b$) vectors
\begin{equation}
{\bf t}_i = \frac{ {\bf r}_{i+1} - {\bf r}_i }{ | {\bf r}_{i+1} - {\bf r}_i |}
\ \ \ \  \& \ \ \ \  {\bf b}_i = \frac{ {\bf t}_{i-1} \times {\bf t}_i }{| {\bf t}_{i-1} \times {\bf t}_i|}
\label{tb}
\end{equation}
With the unit normal vector ${\bf n}_i = {\bf b}_i \times {\bf t}_i$ we have the full orthonormal  Frenet frame at the location of each C$_\alpha$.
The bond angles and torsion angles are
\begin{equation}
\kappa_i \  =   \arccos (  \mathbf t_{i+1} \cdot \mathbf t_i  ) \ \ \ \ \ \& \ \ \ \ \ 
\tau_{i} \  =  \arccos (\mathbf  b_{i+1} \cdot \mathbf b_i )
\label{kt}
\end{equation}
The Frenet framing  describes  the position of all atoms of the protein, in the way how these atoms 
are seen by an imaginary observer who roller-coasts  the backbone along the C$_\alpha$ atoms
with gaze direction always fixed towards the next C$_\alpha$ \cite{dff}.
In Figure 1 we display the statistical angular distribution of the backbone N and C and the side-chain C$_\beta$ atoms
in our PDB data set,   as they are seen by a Frenet frame observer who moves through all the proteins in our data set.
The sphere is  centered at the C$_\alpha$, and its radius coincides with
the length of the (approximatively constant) covalent bond. We take the vector
$\bf t$ that points towards the next C$_\alpha$ to be  in the direction of the positive $z$-axis, so that  with
$\mathbf n$ in the direction of positive $x$-axis we have  a right-handed Cartesian coordinate system. 
We introduce the canonical spherical coordinates,  so that the angle
$\theta \in [0,\pi]$ measures latitude  from the positive $z$-axis   
and the angle $\varphi \in [0,2\pi]$ measures longitude  in a counterclockwise direction from the $x$-axis {\it i.e.} from the
direction of $\mathbf n$ towards that of $\mathbf b$.
%
\begin{figure}[h]
        \centering
                \includegraphics[width=0.85\textwidth]{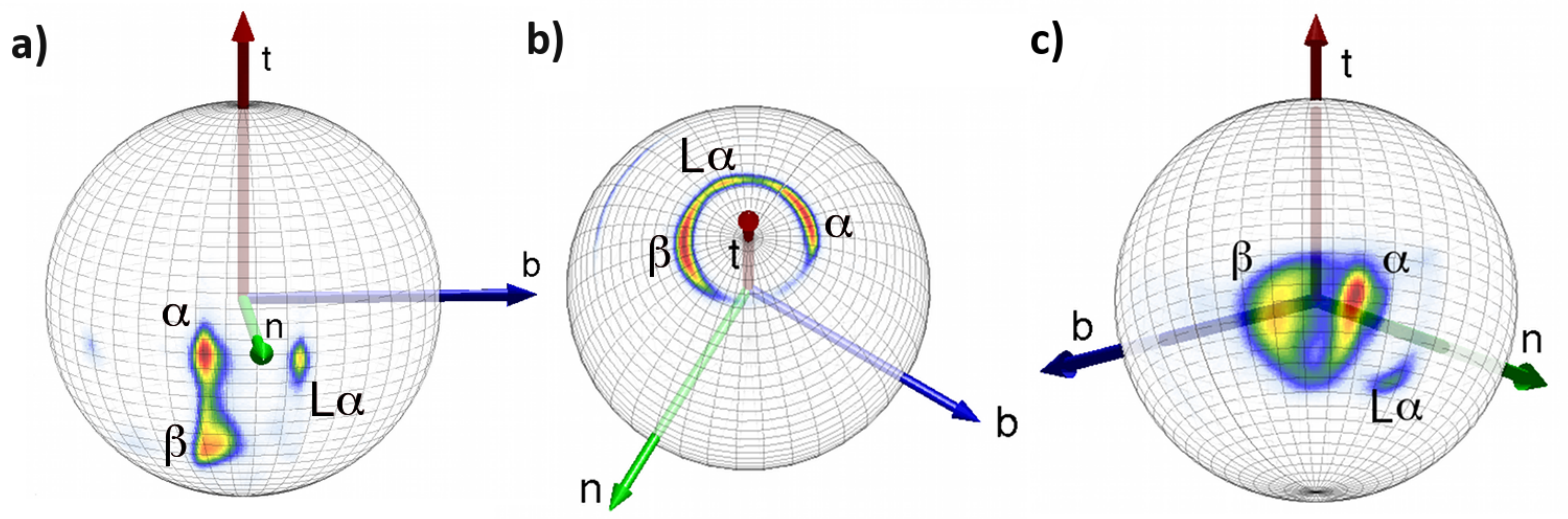}
        \caption{{ \it
     The locations  of a)  backbone N-atoms, b)  backbone C-atoms  c) side-chain C$_\beta$ atoms, 
     as seen by a Frenet frame observer  located at  the C$_\alpha$ carbon 
     at the center  of the sphere.     
     In a) the (smaller) point-like  direction of backbone N atoms corresponds to the  {\tt L}-$\alpha$  Ramachandran region.
     The larger  region forms a segment of the great circle $\varphi \approx -15^{\mathrm o}$. Note that the loops interpolate latitudinally 
     between $\alpha$-helices and $\beta$-sheets.  
      In b) the directions of backbone C form a segment of a small circle around $z$-axis, with $\theta \approx  20^{\mathrm o}$.
      The N and C oscillations become coupled into the horse-shoe shaped nutation of C$_\beta$ as shown in Figure 1c).
       }}
       \label{Figure 1}
\end{figure}

We find it  remarkable that in the Frenet frame coordinate system, the N and  C  oscillations shown in Figures 1 (a) and 1 (b)  
become fully separated into the  locally orthogonal $\theta$ and $\varphi$ directions respectively;  this is not the case in a generic coordinate system.  We also find 
it remarkable that secondary structures such as  $\alpha$-helices, $\beta$-sheets, loops and 
left-handed $\alpha$-regions are all clearly identifiable.
Figure 1 (c) then reveals how the  N and C oscillations become coupled into a horseshoe shaped 
nutation of C$_\beta$. This  nutation is similarly entirely determined by the local secondary structure environment,  in  an equally  systematic manner. 

In Figures 2 (a)-(f) we  plot the   tetrahedral bond angles $\tau_{NC}$,  $\tau_{C\beta}$ and  $\tau_{N\beta}$  separately for  the $\alpha$-helices, $3/10$-helices and  $\beta$-strands; As in figure 1 the loops will continuously interpolate between these regular secondary structures. 
The Figures 2 (a) and 2 (d) clearly reveal that  the  $\tau_{NC}$ angles depend  on the secondary structure in a 
systematic  manner.  But we observe no similar effect in either  
$\tau_{C\beta}$  or  $\tau_{N\beta}$. (The isolated small peak in Figure 2 (b)  and 2 (e) is due to prolines.)

The fact that  {\it only}  $\tau_{NC}$  in Figure 2  displays  systematic 
secondary structure dependence  makes it 
plain and clear that the paradigm hybridized tetrahedral symmetry of the  C$_\alpha$ carbon 
atomic orbitals is broken.  {\it In a folded protein the
covalent tetrahedral structure around C$_\alpha$ is not unique}.   Instead, the backbone secondary structure 
breaks the  ground state tetrahedral symmetry 
in a systematic manner which is fully determined by the  local secondary structure. We note 
that for the  loop regions, the tetrahedron geometry interpolates deterministically 
between those of the adjacent regular secondary structures. There is a 
one-to-one correspondence  between the shapes of the C$_\alpha$ tetrahedra and the way how a protein folds.

On the basis of the present PDB data we are unable to conclude whether the fact that  
the symmetry breaking is visible only in $\tau_{NC}$ reflects a true physical effect, or is simply a consequence  of  the
existing refinement procedures that place all the tension on the $NC$ bond angle. 
We propose that these details of the symmetry breaking could be investigated in the new generation 
ultra-high resolution X-ray experiments.  
%
\begin{figure}[h]
        \centering
                \includegraphics[width=0.85\textwidth]{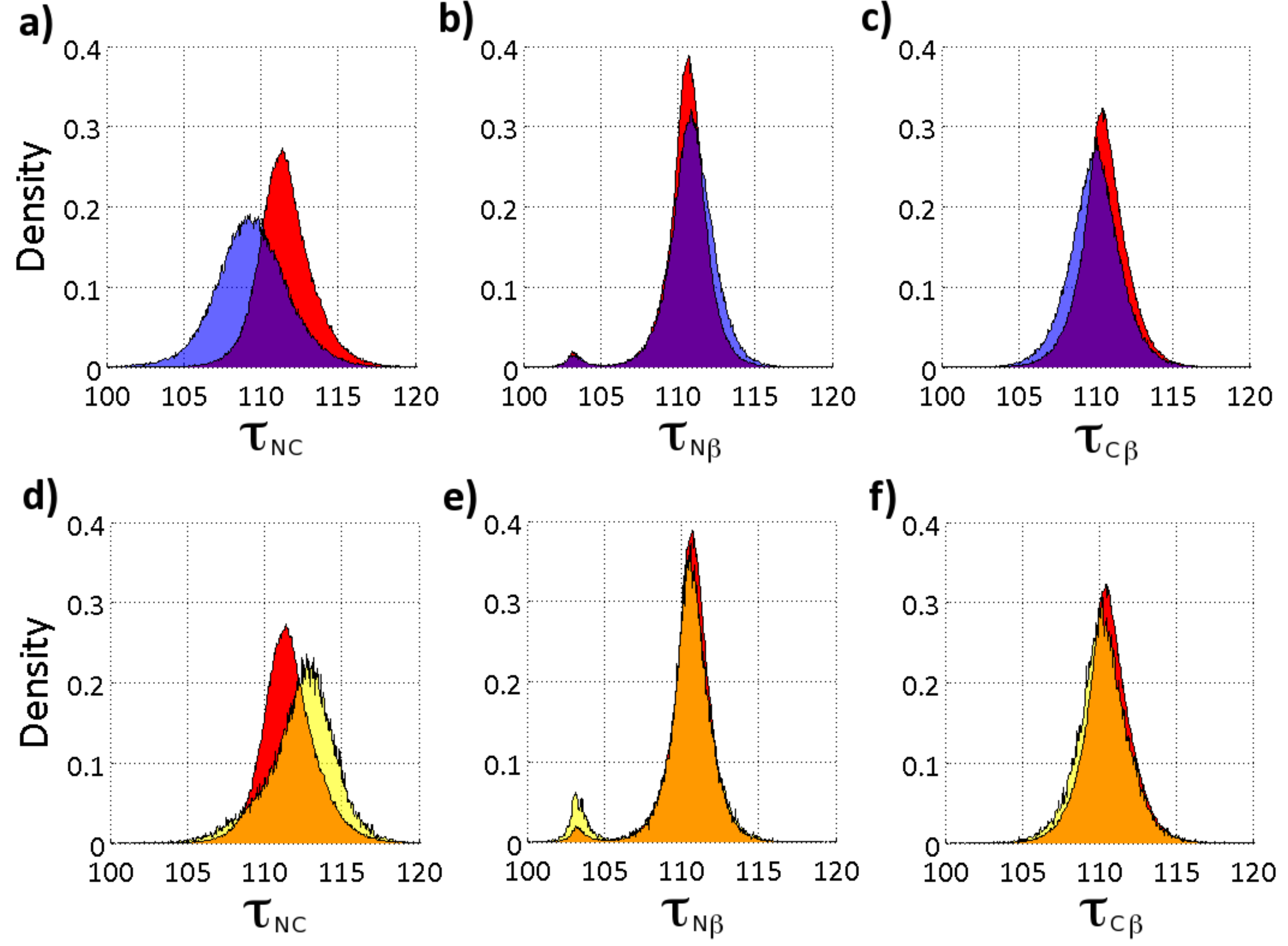}
        \caption{{ \it
     The probability density angular distribution of the $\tau_{NC}$,   $\tau_{N\beta}$ and  $\tau_{C\beta}$ angles  (in degrees) 
     separately for $\beta$-strands (blue) and $\alpha$-helices
      (red) in (a)-(c), and separately for 3/10 helices (yellow) and $\alpha$-helices (again 
      red) in (d)-(f).   The secondary peak in (b) and (e) is proline. }}
       \label{Figure 2}
\end{figure}
%

%
%

{\it Any} molecular dynamics approach \cite{nat} to protein folding is based on a harmonic approximation of
the energy around
the equilibrium values of the bond angles, 
\begin{equation}
E_{bond} = \sum_{\rm bonds} K_\theta (\theta - \theta_0)^2
\label{ba}
\end{equation}
Here the equilibrium values $\theta_0$ are determined using the paradigm 
that the sp3 symmetry of the amino acid remains unbroken and
$\theta_0$ should  have {\it no}  dependence on the eventual
secondary structure environment.  However, Figures 1 and 2 show unequivocally that in actual proteins these
equilibrium values depend on the secondary structure in a systematic manner.  We now proceed to 
investigate how to develop  an energy function that describes this broken symmetry.  Our starting point is
the backbone energy function of  \cite{oma2},  there it has been shown how the collapsed  PDB proteins
can be described with experimental B-factor accuracy in terms of 
soliton solutions to a generalized discrete nonlinear Schr\"odinger (DNLS) equation.  Indeed, 
the soliton solutions  of DNLS equation share a remarkable history with protein research \cite{dnls}, 
they were first introduced by Davydov to describe the propagation of energy along $\alpha$-helices \cite{davy}.
Mathematically the DNLS equation is {\it integrable}, there is an infinite hierarchy of conserved quantities  
\cite{fadde}.  Explicitely the backbone energy function is \cite{oma2}
\begin{equation}
E = - \sum\limits_{i=1}^{N-1}  2\, \kappa_{i+1} \kappa_i  + \sum\limits_{i=1}^N
\left\{ 2 \kappa_i^2 + q\cdot (\kappa_i^2 - m^2)^2 
+ \frac{d_\tau}{2} \, \kappa_i^2 \tau_i^2  -  b_\tau \kappa_i^2  \tau_i - a_\tau  \tau_i   +  \frac{c_\tau}{2}  \tau^2_i 
\right\} 
\label{E1}
\end{equation}
Here the first sum together with the three first terms in the second sum comprise the energy function of  the conventional
DNLS equation, when expressed 
in the standard  
Hasimoto variables  of fluid mechanics, see \cite{dff}- \cite{oma2}  for full details. 
The fourth ($b_\tau$) and fifth ($a_\tau$) terms are the 
{\it only} two lower order nontrivial
conserved quantities that appear in the integrable  DNLS hierarchy prior to the energy. These are  the momentum and the helicity, respectively.  
The last ($c_\tau$) term 
is the standard Proca mass term that we add for completeness, it could be ignored with only a minor effect on accuracy.  
Note in particular that any term odd in the $\kappa_i$ is excluded by a global $\mathbb Z_2$ symmetry \cite{dff}.
 We also note that the next, higher order conserved quantity 
in the DNLS hierarchy is the energy function of the  modified KdV equation \cite{fadde}. It could be added, but there is no point since with 
the present energy function we already reach the experimental B-factor accuracy.

The remarkable property of (\ref{E1}) is  that the torsion angle $\tau_i$ appears at most quadratically so that it can be eliminated explicitly by using
equations of motion. The values of the torsion angle and consequently the entire C$_\alpha$ backbone geometry becomes  then fully
determined by bond angle soliton solutions of a generalized DNLS equation \cite{oma2}.

The Figures 1 (a) and (b) reveal that the N and C atoms oscillate independently, in the latitudinal and longitudinal directions respectively.
Consequently the ensuing contributions to the protein energy function should also  be independent and 
depend only on the respective
angular variables.  Together these two independent contributions should then combine
into the C$_\beta$ nutation of Figure 1 (c). 

Combining standard universality arguments   with  the
spirit of the harmonic approximation (\ref{ba}),  we  
propose that the latitudinal and longitudinal contributions to protein energy only involve the 
two lower order conserved quantities in the integrable DNLS hierarchy and the Proca mass.
This fixes the ensuing contributions uniquely,
\begin{equation}
E_\theta  =  \sum\limits_{i=1}^N
\left\{  \frac{d_\theta}{2} \, \kappa_i^2 \theta_i^2  -  b_\theta \kappa_i^2  \theta_i - a_\theta \theta_i   +  \frac{c_\theta}{2}  \theta^2_i 
\right\}   
\label{E2a}
\end{equation}
\begin{equation}
E_\varphi  =  \sum\limits_{i=1}^N
\left\{  \frac{d_\varphi}{2} \, \kappa_i^2 \varphi_i^2  -  b_\varphi \kappa_i^2  \varphi_i - a_\varphi \varphi_i   +  \frac{c_\varphi}{2}  \varphi^2_i 
\right\} 
\label{E2b}
\end{equation}
Accordingly the spherical angles  ($\theta_i, \varphi_i$) are  fully  determined  by the profile of the backbone bond angles $\kappa_i$ 
and the {\it global} parameters that are specific  only to a given super-secondary structure.  In particular, the tetrahedral symmetry breaking
becomes driven by the degenerate ground state structure of the DNLS equation.  (We note   that  the $\theta_i$ and $\varphi_i$ variables 
are coupled to each other only indirectly, by the bond angles $\kappa_i$. In particular, the long range interactions 
that are necessary for describing a collapsed protein are entirely due to the non-local character of the
DNLS solitons.)

Since (\ref{E2a}), (\ref{E2b})  involves both latitudinal and longitudinal angles,  the classical solutions of  
 (\ref{E1})-(\ref{E2b})  can be utilized to describe  both the backbone  C$_\alpha$ and the side-chain C$_\beta$ atoms
 in a  folded protein. As an example we inspect 
the chicken villin headpiece subdomain HP35 with PDB code 1YRF. This is a naturally existing 35-residue 
protein with three $\alpha$-helices separated from each other by two loops. The villin continues to  be  subject 
to very extensive studies both experimentally \cite{meng}-\cite{wic} and {\it in silico} \cite{pande2}-\cite{sha}, and  
\cite{sha} reports on a molecular dynamics construction of folded villin with overall backbone RMSD accuracy around one \.Angstr\"om.
Since the force fields in  \cite{pande2}-\cite{sha}  utilize the paradigm concept  of unbroken C$_\alpha$ tetrahedral symmetry, 
the accuracy of in particular  \cite{sha}  can be adopted as a good measure of the symmetry breaking effect.

We first solve for  the classical equations of motion for $\kappa_i$ and $\tau_i$ from  (\ref{E1}), and then
construct the remaining angular variables from (\ref{E2a}), (\ref{E2b}) in terms of $\kappa_i$.  We use the iterative algorithm and procedure that has been
described in 
\cite{her}, \cite{oma2}. Using the parameters in Table 1 we reach an overall RMSD accuracy  0.39 \.A for the combined   
C$_\alpha$-C$_\beta$ configuration which is in line with the experimental B-factor accuracy; 
see Figure 3 that displays our solution in comparison with the PDB configuration. 
%
\begin{figure}[b]
        \centering
                \includegraphics[width=0.65\textwidth]{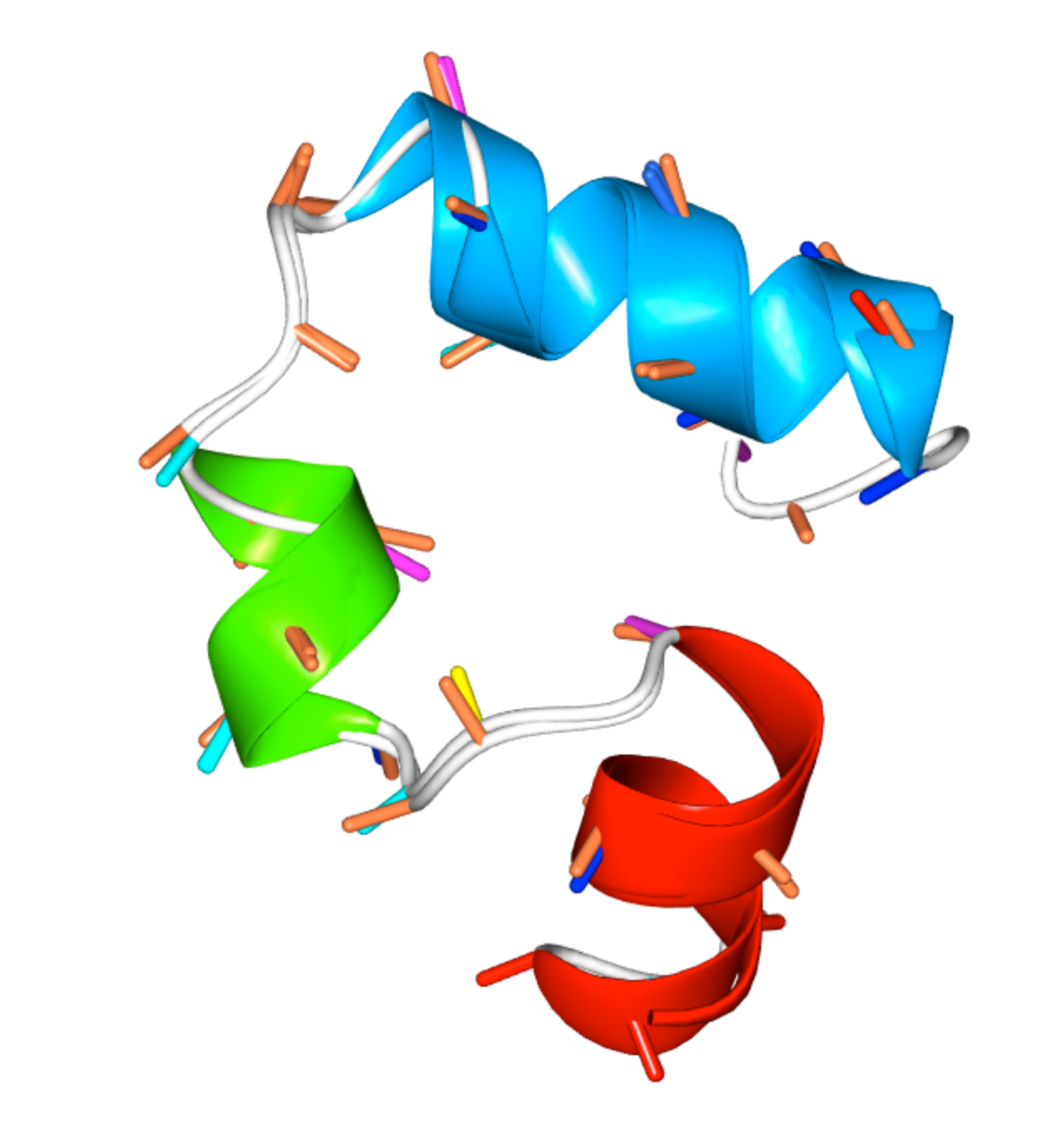}
        \caption{{ \it
    A cartoon comparison of HP35 with our soliton solution summarized in Table 1. The combined C$_\alpha$ and C$_\beta$
    root-mean-square distance is 0.39 \.A which is in line with  the experimental B-factor accuracy.  }}
       \label{Figure 3}
\end{figure}
%
%
%

%

{\small
\begin{table}[b]
       \centering
\begin{tabular}{|c|c|c|}
\hline  parameter & soliton-1 & soliton-2    \tabularnewline
\hline \hline
$q_1$ & 0.459712  &   0.995867 \tabularnewline
\hline
$q_2$   & 4.5533320 &  9.408796  \tabularnewline
\hline
$m_1$ & 1.504535 & 1.550322  \tabularnewline
\hline
$m_2$  & 1.512836  &  1.535081   \tabularnewline
\hline
$a_\tau$ & 9.5752137e-9 & 7.840467e-6   \tabularnewline
\hline
$b_\tau$ & -676965e-11 &  -4.973244e-9  \tabularnewline
\hline
$c_\tau$ & 4.875744e-9 &  4.2733696e-6  \tabularnewline
\hline
$d_\tau$ & -2.917129e-9 &  -2.431388e-6   \tabularnewline
\hline
 $a_\theta$ &1.514770  &  1.322495  \tabularnewline
\hline
$b_\theta$ & -0.0017952 & -0.018619  \tabularnewline
\hline
$d_\theta$ &  0.0420877 &  6.930946e-8  \tabularnewline
\hline
$a_\varphi$ & 0.544859 &   0.3594184 \tabularnewline
\hline
$b_\varphi$ & 5.66111e-5  & 3.83253e-4   \tabularnewline
\hline
$d_\varphi$ & -0.1845828  &  -0.226012  \tabularnewline
\hline  \hline
RMSD (\.A) & 0.38 & 0.32 \tabularnewline
\hline
\end{tabular}
\caption{\small
\it Parameter values for the two-soliton solution that describes the two
loops of  1YRF with 0.39\.A accuracy for both
C$_\alpha$ and C$_\beta$ atoms.  The displayed  RMSD values are for the
individual solitons. 
The soliton-1 is located at  Glu-45 - Phe-58 and the soliton-2 is
located at Phe-58 - Lys-73.
We utilize scale invariance to set all $c_\theta = c_\varphi = 1$. Notice that the result is
sensitive to the accuracy of parameters. This is because a folded protein is a piecewise linear
polygonal chain with a positive Liapunov exponent.
 }
       \label{tab:para}
\end{table}
}

Symmetry breaking is a fundamentally important physical phenomenon,  often 
intimately related to structure formation. Here we have shown that a protein backbone breaks the tetrahedral 
symmetry of the sp3 hybridized  C$_\alpha$ covalent bonds,  in a  manner that is  entirely
determined by the local secondary structure.  We have also presented  a simple energy function that accounts for the symmetry breaking, to compute the C$_\beta$ nutation trajectories of the HP35 villin with experimental
B-factor accuracy.  Our observation is based on the available high precision  PDB data, 
consequently detailed analysis of our symmetry breaking is experimentally feasible.  
Our observations should have wide applicability 
in the development of future refinement  tools, and for  constructing accurate  theoretical and computational methods for investigating   protein folding dynamics  and structure. Indeed, the direct 
relation between the symmetry breaking and the protein fold geometry suggests
that our broken symmetry is intimately connected to the underpinnings of protein folding and thus 
with the origin of life.

\end{document}